# DISCOVERING USERS TOPIC OF INTEREST FROM TWEET

Muhammad Kamal Hossen[1], Md. Ali Faiad[1], Md. Shahnur Azad Chowdhury[2] and Md. Sajjatul Islam[3]

[1]Department of Computer Science and Engineering, CUET, Chittagong, Bangladesh
[2]Department of Business Administration, IIUC, Chittagong, Bangladesh
[3]Department of Computer Science and Engineering, CIU, Chittagong, Bangladesh

## ABSTRACT

*Nowadays social media has become one of the largest gatherings of people in online. There are many ways for the industries to promote their products to the public through advertising. The variety of advertisement is increasing dramatically. Businessmen are so much dependent on the advertisement that significantly it really brought out success in the market and hence practiced by major industries. Thus, companies are trying hard to draw the attention of customers on social networks through online advertisement. One of the most popular social media is Twitter which is popular for short text sharing named 'Tweet'. People here create their profile with basic information. To ensure the advertisements are shown to relative people, Twitter targets people based on language, gender, interest, follower, device, behaviour, tailored audiences, keyword, and geography targeting. Twitter generates interest sets based on their activities on Twitter. What our framework does is that it determines the topic of interest from a given list of Tweets if it has any. This process is called Entity Intersect Categorizing Value (EICV). Each category topic generates a set of words or phrases related to that topic. An entity set is created from processing tweets by keyword generation and Twitters data using Twitter API. Value of entities is matched with the set of categories. If they cross a threshold value, it results in the category which matched the desired interest category. For smaller amounts of data sizes, the results show that our framework performs with higher accuracy rate.*

## 1. INTRODUCTION

General people now gather in popular social media, such as Facebook, Twitter, Instagram, LinkedIn, WhatsApp, etc. The number of people gathered here is big and can be compared to a big market for industries. This is an important fact but also is very tough to know the taste and interests of each people for advertising. User-generated content is the lifeblood of social media [1].They stores information about users related to their age, gender, interests, location, something they shared, etc. This stored information allows advertisers to target specifically a particular group to show their advertisements. The advertisers can thereby analyse the stored information to categorize the interests of that individual group. This is an advantage for the advertisers and that specific groups of people will only be shown the advertisements according to their taste. Social media like Twitter also follow this method. This is why Twitter is so famous for advertising in online.To advertise on Twitter, there are promoter tweets, trends, promoted accounts and other things that show up on users' newsfeeds [2].Most of the Twitter users use English language and so, nearly half of the tweets are in the English language. This is why the English language plays a significant role in analysing tweets. Our framework can analyse only English tweets.

As tweet does not necessarily maintain the grammatical rules, direct language processing algorithms that maintain grammatical rules are not implemented in this framework. Furthermore, discovering interest set is one very much useful for advertising on Twitter. Every advertiser has a list of Twitter users who have been on their advertisement's list and bought a product. This list contains their usernames and advertisers follow them. From this given username, we can retrieve their tweets that can be very much advantageous for knowing their interests in different topics for retargeting. All these motivated us to design a system that is not only grammar independent analyser but also useful for the advertisers.

Our work is done using some free API (Application Program Interface). For the time limit for a given request, we could only process one tweet at a time and discover interested topics. Moreover, we have some 'false-positive' issues. Other than that, it has greater accuracy rate and is implemented for the testing condition.

The remainder of this paper is as follows, section 2possesses some related works. Section 3 formally presents the problem. Section 4 describes the design of the framework. Section 5 shows the experimental results and observations where accuracy is calculated. Lastly, we concluded the paper in section 6 where we also suggested some future recommendations.

## 2. RELATED WORKS

Before starting our work, we followed some other works that led us to this project. These works show the targeted advertisings' effectiveness that advertises leading to more profits for the products and how free API is used for social data collection. Chan Mei Lee et al. mentioned that since using the traditional survey method will limit the coverage of different demographic people and increase the cost, hence the future researcher is highly encouraged to perform an online survey as the targeted respondents of this research are internet users [3].

The study of Billy Bai et al. attempted to explore the marketing effectiveness of two different social media sites (Facebook and Twitter). This study also indicated that different social media sites demonstrate the same marketing effectiveness [4].Gupta et al. promoted posts that provide businesses with the ability to push a post out to not only fans but likers of fans, increasing the reach dramatically. Offers allow businesses to present advertised offers which spread virally as people accept the offer acting as brand advocates for one's brand by pushing this to their friends. This is really only the tip of the Iceberg. Social media is projected to ramp up its revenue-generating activities [5].

A study was conducted by B. Pikas et al. to test Americans perception of online advertising on the popular web pages such as Facebook, YouTube, and Twitter [6].The framework proposed by Kenny Ho et al. helps to identify the most effective and efficient spammer detection features, evaluate the impact of using different numbers of recent tweets, and therefore, obtaining a faster and more accurate classifier model [7].

According to Waring et al., given an advertising category, it can be estimated the most promising areas to be selected for the placement of an advertisement that can maximize its targeted effectiveness [8].Uchiya et al. specified that when using Twitter, it is difficult for users to find appropriate users to follow among the vastly numerous users [9]. To solve this problem, we propose a follow-user recommendation system on Twitter based on an interest domain. Moreover, we demonstrate the effectiveness of the proposed method through experimentation.

Liu et al. proposed a framework through which important research themes and concepts are synthesized to provide IS researchers with an overview of this research area and to identify those topics where much research has already been done and those topics where more research is needed [10]. For click-through prediction, X. He et al. stated that the task of click-through prediction for online advertising at Twitter is related to the practice in the social media industry, including online advertising on Facebook, LinkedIn, etc. The most relevant work in literature is a recent study of predicting clicks on Facebook ads [11].McCorkle et al.presented details for implementing a Twitter project in a digital marketing course as a means of addressing these gaps. Student's feedback on this project indicated that their skills in social media marketing, digital marketing, and personal branding were improved. Suggestions for use of Twitter in other courses are also presented [12].

## 3. PROBLEM DEFINITION

EICV stands for Entity Intersect Categorized Value. It mainly finds the interested topic from tweet for a given set of interests. The equations for EICV are described below. Here,

$W_n$ = set of words,

$R_n$ = set of redundant words and phrases,

$K_n$ = set of retrieved words from knowledge source for Entities,

$T_n$ = set of topics,

$T_{tag[n]}$ = set of tags for each topic $T_n$.

If an element $W_n$ is related to a topic in $T_n$, then it creates,

$$K_n = F_{twitter}(W_n - R_n) \qquad (1)$$

where, $F_{twitter}$ is a function that generates entities from knowledge source (for our case, it is Twitter search API). The equation for intersection value is

$$V_i = number\ of\ (K_n \cap T_{tag[n]}) \qquad (2)$$

For each topic from $T_n$, it generates an intersection value. We will set a threshold value $T_v$. The intersection values greater than the threshold value are our desired topics.

## 4. EICV - SYSTEM ARCHITECTURE

The architecture of the system comprises four basic modules. These modules are:

- ➢ Topic and tag input module
- ➢ Tag generator module
- ➢ Tweet input module
- ➢ Interest generator module

The first module takes the advertisement topics and the tags for each advertisement topic as input. The second module generates related words and phrases. This words and phrases are normally ambiguous. So, after detecting the ambiguous words and phrases, we remove as much ambiguity as possible. In the third module, Twitter data can be inputted. We can input Tweets directly into the system or enter a username to retrieve the last tweet of that account. Lastly (in the interest generator module), the matched topics to the tweet are generated.

## 4.1. Topic and Tag Input Module

### 4.1.1. Topic Input

Firstly, the topics to be advertised are defined. There is a list of advertisement topics in the Twitter advertisement section. In our system, we input the list of topics in the *csv* file. Algorithm 1 states the process of inputting topics and tags.

**Algorithm 1.** Input topics and tags

 i. Start
 ii. Input topics into the *csv* file
 iii. Create a *csv* file for each topic
 iv. Input tags into the correspondent *csv* files
 v. End

For only simplicity of the research purposes, we took 10 topics that are randomly selected. But the system is designed to be extendable. That means, we can take as many topics as we like. These topics can be deleted too. This is the set of topics, $T_n$.

### 4.1.2. Tag Input

After topics being set, a *csv* (comma separated values) file is created for each topic. This type of file contains different values separated by a comma. Tags are words or phrases that are related to this topic. For each topic, we can enter tags in the corresponding *csv* file. For example, if we have a topic named "Football", a file will be created named "football.csv" and we can put the tags on there.

## 4.2. Tag Generator Module

Algorithm 2 states the process of generating tags using d*atamuse* API and elimination of ambiguous tags.

**Algorithm 2.** Generating tags using d*atamuse* API

 i. Start
 ii. Take topic names from the *csv* file
 iii. Generate tags for each topic into the correspondent *csv* file
 iv. Find the ambiguous words and phrases
 v. Delete the ambiguous words and phrases
 vi. End

### 4.2.1 Datamuse API

As tags selected by human are not enough for this research, we took help of API (Application Program Interface). These are the tools that enable us to use any external application through interface either on free of any service cost or on paid terms. We used a free API for this purpose named *datamuse* API. It provides a lot of features including finding tags for any topic. This is important to have as many tags as we can get since it results in higher accuracy.

### 4.2.2 Ambiguity

As being free API, this does not generate tags which denote that topic only. Moreover, there are a lot of words generated that do not have the quality to be tagged. These words are called ambiguous words. We made a way to find these ambiguous words and show them to the trainer. A list of these ambiguous words is shown for each topic.

### 4.2.3 Eliminating Ambiguity

To be able to make this system working, we have to train it first. There is a list of ambiguous words for each topic. The tags that have ambiguity are selected and removed. But, there are a lot of tags that still remain in more than one topic's tag list. For example, "player" is a tag that is generated by both "Football" and "Cricket". This is not ambiguous as the tag "player" signifies for both the topic. After eliminating all ambiguities, this module is ready.

## 4.3 Twitter Input Module

In this work, we will only analyse the tweets. Tweets are generally 140 characters long (from November 5, 2017). To check the desired interest on topics by analysing tweets we have shared, it is needed to enter tweets into the system. A tweet can be entered into the system by two ways.

These are:

i. Enter tweet directly into the system,
ii. Enter Twitter's username to retrieve the last tweet of that user.

Algorithm 3 states how tweets can be inputted into the system.

**Algorithm 3.** Twitter data input

i. Start
ii. Input tweets directly into the system and go to (v)
iii. Otherwise, take Twitter's username as input
iv. Collect the last tweet using the given username
v. End

### 4.3.1 Direct Tweet

The tweet that we want to analyse can be entered directly into the system.

### 4.3.2 Twitter's Username

We can also enter any valid username. Luckily, Twitter has their own API named *Twitter* API, free of any charge that allows us to collect all the tweets with details in different formats. *Twitter* APIis used to get only the last tweet of a user only for simplification of the procedure. This system can be extended to work with all the tweets of a particular user.

In both ways, the system gets the tweet which is ready to be analysed.

## 4.4 Interest Generator Module

Interest generator module is the key part of this system. It takes tweets and tags as input and returns set of interests as output. The whole procedure is done in several stages. These stages sum up the core functionality of the system. The interest generator module can be divided into six stages to be exact.

They are:

- Redundant words remove
- Keyword generation
- Gather knowledge according to keywords
- Ranking
- Find intersect values
- Generate topics

Algorithm 4 states the process of generating tags using *datamuse* API and elimination of ambiguous tags.

**Algorithm 4.** Generating topics' list to be advertised

 i. Start
 ii. Eliminate unnecessary words and phrases
 iii. Generate keywords
 iv. Search Twitter with keywords
 v. Retrieve results
 vi. Rank frequent words
 vii. Match most frequent words to the tags
 viii. Find tags that are most common with the topics
 ix. Display the topics
 x. Display error rates
 xi. End

### 4.4.1 Redundant Words Remove

Tweets do not necessarily maintain the grammatical rules as they are really hard to be analysed by natural language processing. Instead, this method divides the tweets into words that are called entities. This system is only being tested for the English language with the purpose of testing. The entities are mostly full of redundant words, such as "lol", "goooaaal", "suppperb", URLs, etc. Now, we have to find $K_n$ that defines the set of words extracted from the knowledge source. The redundant words, for this reason, have to be deducted. So, initially, all the redundant words have been removed.

In this stage, the value of ($W_n$-$R_n$) has been determined where $W_n$ denotes set of words or entities of the tweet and $R_n$ denotes set of redundant words.

All the URLs have been removed as the tweets are only taken as input. They are removed through Regular Expression (RE). Then, a set of words is created in a *csv* file named "redword.csv". These words are the redundant words which are not necessary for the analysis that can be denoted by $R_{nr}$. So, deleting all the redundant words, it remains ($W_n$ - $R_{nr}$) words where,

$$R_n = R_{nr} + R_{nk} \qquad (3)$$

Here, $R_{nk}$ are the words we could not delete.

### 4.4.2 Keyword Generation

Since the numbers of words which are not necessary to be the keywords are numerous, we took help of a free API supported by "cortical.io". After deleting all the redundant words, the system sends the ($W_n$-$R_{nr}$) words to the website. It deletes all the remaining unnecessary words $R_{nk}$ and we got back,

$$F_{input} = W_n - R_n \qquad (4)$$

Since this is a time-consuming process, a process is utilized to add words, $R_{nk}$ to the $R_{nr}$ so that after a long period of time, our own $R_{nr}$ set will be enough for finding the keywords. The keywords extraction API will no more be needed.

After this stage, we have the set of keywords ($W_n$-$R_n$) which are prepared to be inputted into the next stage.

### 4.4.3 Gather Knowledge According to Keywords

After having the keywords, we will search into the Twitter with those keywords. Twitter supporting free API has an option to search tweets on the basis of keywords matching to them. There are two categories of tweets we can find. They are:

- Popular tweets
- Recent tweets

Twitter restricted the users to have popular tweets upto 20 tweets and recent tweets up to 100 tweets. We retrieved all the tweets and made up an array of all the words of these tweets.

### 4.4.4 Ranking

After collecting all the words of the retrieved tweets, the array words' frequency is counted. Then, the words are sorted in descending order depending on their frequency of occurrence. Gathering knowledge according to keywords and ranking, together it denotes the function $F_{twitter}()$. This function takes the argument of entities ($W_n$-$R_n$). After frequency counting is done, we choose those words whose frequency is more than three at least and take a certain number of words. This set of words is $K_n$.

### 4.4.5 Find Intersect value

Now, we have $K_n$ and $T_{tag[n]}$. Both of these are set of words or phrases. Then the system finds a number for each topic. This number is determined by the number of words in the set $K_n$ matches with the number of words or phrase of $T_{tag[n]}$.

Suppose, we have a topic "Football" that has 200tags denoted by $T_{tag[football]}$. For an input tweet, supposeit is found 20 words denoted by $K_{tweet}$. If 10 words are matched, then the intersection value,

$$V_{football} = number\ of\ (K_{football} \cap T_{tag[football]}) = 10 \qquad (5)$$

So, for each topic $T_n$, we will get a value of $V_n$. Finally, we have the intersection values for each topic.

### 4.4.6 Generate Topics

From the intersection value, we will first find the maximum value $T_{max}$. This is the primary topic of advertisement for that particular tweet. Since each tweet can lead us to more than one topic also, we set another process to get other interest topics. We set a threshold value,

$$T_v = 3 * T_{max}/4 \qquad (6)$$

This threshold value is changeable. Initially, we set it on this which leads us least error rate. All the topics which maintain $V_n > T_v$ for each topic are our topics of advertisement. If the value of $T_{max}$ is very low, the number of matched topics $V_n$ become high. So, we added an extra feature that the tweet will only generate topics when,

$$T_{max} > Minimum\ Value \qquad (7)$$

By default, this Minimum Value is also changeable according to practices. Finally, this system returns the set of topics to be advertised for a given set of topics to be advertised and a given tweet.

## 5. EVALUATION OF RESULTS

The framework has been tested for over 1378 tweets or usernames (till December 15, 2017). As for the experimental review, we specified our system with only ten particular topics which are: football, cricket, golf, baseball, movie, book, food, politics, drink, and science. We have also tested by increasing the number of topics to 25 topics which really did not affect the accuracy of the system. Some of the test cases are shown in Table 1.

Table 1. The generated interest of topics for the proposed system.

| **Tweets** | **Matched Topic** |
|---|---|
| I like Ronaldo | Football |
| I like Harry Potter | Movie |
| How good Shahrukh Khan is!!1! | Movie |
| This time I am not going to miss pizza | Food |
| I am frustrated | No topic matched |
| Sachin is my idol | Cricket |

The tweet "I am frustrated" did not result in any tweet because it did not match with any of the advertisement topics which are true. This clarifies that TweetAd (our proposed system) generates all the interests of topics correctly from only tweets for this set of results.

For calculating accuracy, we used the following equation:

$$Accuracy\ rate = \frac{Number\ of\ accurate\ topic\ generation * 100}{Number\ of\ test\ cases} \quad (8)$$

It has correctly discovered 1282 times among 1378 tweets or usernames and could not detect correctly 96 times. So, the accuracy rate is 93.03%. Again to be mentioned, let we have a tweet with an entity "Ronaldo". If we test this on the framework, most of the times it displayed only one topic "Football" which is considered as correct, $2^{nd}$ most topic generation was both "Football" and "Cricket" which is also considered as correct, and $3^{rd}$ most frequent result was "Cricket" and "Football" which is also considered as correct. We considered a topic being guessed correct if the topic is at most of the $2^{nd}$ rank and that did not result in any other topic rather than it being a game. If one of the topics is "Politics", we would not consider that as an accurate topic generation.

## 6. CONCLUSIONS

Advertising is one of the hottest topics in recent days in social media. Researchers are working hard for a long time to pinpoint Twitter users' interest in an advertisement topic. The performance of the existing system is good. Still, a lot of targeting options are missed through the crack of the system. The proposed system can be used in two ways. In one way, implementing it in Twitter, it would have more data for analysing which might result in greater accuracy. In another way, it can be used as an external application for the advertisers who can determine the people interested in their products with more accuracy alongside with Twitter advertisement system.

As every system has some limitations, this system also has that are stated below:

- This system detects wrong topics in a rare case as the resources were limited.
- This system will only work for the English language.
- There are some cases where topics are not generated, though it surely has a topic to be generated.
- If any tags are given as input, this system can only retrieve the last tweet only
- Using keyword generation API supported by cortical.io is not a great solution which is time-consuming.
- As we used free API, the number of requests is limited. It will work unlimited on paid service.

As this is only one leap toward a development, there are a lot of works can be done to improve this system or leading to another great system. Future systems can have features for other languages also. Increasing the data source, the tags generated for each topic will be more accurate. Future systems can determine interest set for all tweets for a given user. Lastly, our suggestion will be to make a complete advertisement system along-side with the existing method which will be perfect for commercial uses.

**AUTHORS**

**Muhammad Kamal Hossen** has received his B. Sc. and M. Sc. in Computer Science & Engineering (CSE) degrees from the department of Computer Science & Engineering of Chittagong University of Engineering & Technology (CUET), Bangladesh in 2005 and 2015 respectively. He is now pursuing his Ph. D. degree in CSE from the same university. Since 2006, he has been serving as a faculty member in the Department of CSE, CUET. His research interests include digital image processing, cryptography, steganography, and pattern recognition, data mining.

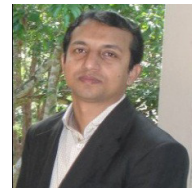

**Md. Ali Faiad** received B. Sc. in Computer Science & Engineering degree from Department of Computer Science and Engineering of CUET), Bangladesh in 2017. His research interests include data privacy and data mining, distributed systems and cloud computing, big data management.

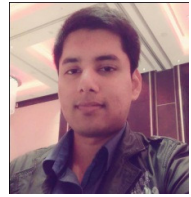

**Md. Shahnur Azad Chowdhury** received B. Sc. in Computer Science and Engineering from International Islamic University Chittagong (IIUC), Bangladesh in 2003 and M. Sc. in 2012 from Daffodil Int'l University, Dhaka. He has been serving the IIUC for the last fourteen years as a lecturer, assistant professor and associate professor respectively. His areas of research interest are natural language processing, data mining, IoT.

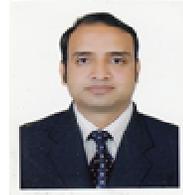

**Md. Sajjatul Islam** has received his B. Sc. in Computer Science & Engineering (CSE) degree from the department of CSE of Chittagong University of Engineering & Technology (CUET), Bangladesh in 2005 and and Master of Science in Computing from University Of Wales (UoW), United Kingdom in 2011. He is pursuing his Ph. D. degree in CSE from CUET. He is now serving as an assistant professor in the Department of CSE of Chittagong Independent University (CIU), Bangladesh. His research interests include big data analysis, cloud computing.
.

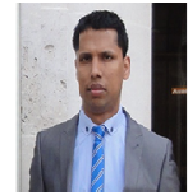